\documentstyle[12pt,aasms4,graphicx]{article}

\def\ltsima{$\; \buildrel < \over \sim \;$}
\def\lsim{\lower.5ex\hbox{\ltsima}}
\def\gtsima{$\; \buildrel > \over \sim \;$}
\def\gsim{\lower.5ex\hbox{\gtsima}}

\begin{document}
\begin{center}

\vspace{1cm}

{\bf \Large GRBs Neutrinos as a Tool to Explore Quantum Gravity
induced Lorentz Violation}

{\bf \large Uri Jacob$^\clubsuit$ and Tsvi Piran$^\spadesuit$}

\noindent {Racah Institute for Physics, The Hebrew University,
Jerusalem, 91904, Israel}
 \end{center}
$^\clubsuit$ {uriyada@phys.huji.ac.il}\hfil\break
$^\spadesuit$ tsvi@phys.huji.ac.il

\begin{abstract}{

Lorentz Invariance Violation (LIV) arises in various
quantum-gravity theories [\cite{QG1,QG2}]. As the typical energy
for quantum gravity is the Planck mass, $M_{pl}$, LIV will, most
likely,  be manifested at very high energies that are not
accessible on Earth in the foreseeable future. One has to turn to
astronomical observations [\cite{QG2}-\cite{SS05}].
Time of flight measurement from different astronomical sources set
current limits on the energy scale of possible LIV to $> 0.01
M_{pl}$ (for n=1 models) and $> 10^{-9} M_{pl}$ (for n=2).
According to current models Gamma-Ray Bursts (GRBs) are
accompanied by  bursts of high energy ($\gsim 100$TeV) neutrinos
[\cite{WB,Vietri}]. At this energy range the background level of
currently constructed neutrino detectors is so low that a
detection of a single neutrino from the direction of a GRB months
or even years after the burst would imply an association of the
neutrino with the burst and will establish a measurement of a time
of flight delay. Such  time of flight measurements provide the
best way to observe (or set limits) on LIV. Detection of a single
GRB neutrino would open a new window on LIV and
 would  improve current limits by many orders
of magnitude. } \end{abstract}

History tells us that symmetries, observed over a large range of
parameters and believed to be fundamental properties of our
physical world, may lose their significance later when
observations are made over a larger range of parameters and when a
new physical understanding arises. Such apparent symmetries often
emerge as leading order approximations of more complex symmetries,
found to describe more accurately the larger range of
observations.

Lorentz invariance might be such a symmetry. A growing number of
speculations suggest that Lorentz invariance might be violated or
deformed at very high energies. These speculations arise in some
theories of Quantum Gravity in which Lorentz violation appears
(see e.g. [\cite{QG1,QG2}] for a review). They are motivated,
independently,  by attempt [\cite{GM98}-\cite{LIDparadox}]
to resolve the GZK paradox [\cite{G66,ZK}]: the arrival on Earth
of ultra high energy cosmic rays at energies above the expected
GZK threshold. LIV would also explain the observations of $20$TeV
photons from Mk 501 (a BL-Lac object at distance of 150 Mpc)
[\cite{TeVgamma-LID1,TeVgamma-LID2,LIDparadox}]. These photons
should have been annihilated via pair creation  with the IR
background.

We consider here a simple phenomenological approach
[\cite{LIDparadox}] for LIV with a symmetry breaking energy scale
$\xi M_{pl}$.
Taking into account only the leading order correction we expect,
for particles with $E \ll \xi M_{pl}$, a generic  approximate
dispersion relation:
\begin{equation}\label{LIDform}
E^2-p^2c^2-m^2c^4\simeq\pm E^2\left(\frac{E}{\xi_n
E_{pl}}\right)^n.
\end{equation}
We consider a single $\xi$ for all particles as such correction is
the simplest and it arises naturally from any theory in which the
modification arises from a small scale structure of spacetime.
Assuming the standard relation $v=dE/dp$ holds the $+$
($-$)\footnote{The increase in the reactions' thresholds that
resolve the GZK and the TeV photon paradoxes requires the $-$
sign.} sign accounts for superluminal (infraluminal) motion.

The most generic attempts to constrain the LIV scale are based on
the energy dependent delay\footnote{ Loosely speaking we use the
term delay to imply both a delay  (corresponding to a $-$ sign in
Eq. \ref{LIDform}, and an early arrival (corresponding to  a $+$
sign).} in arrival of high energy particles [\cite{QG2}]. As no
delays have been observed in GRB [\cite{Ellis}-\cite{EMN06}],
flaring AGN [\cite{AGNs}], or TeV emission from the Crab pulsar
 [\cite{Crab}]  we have only lower bounds: $\xi_1 \gsim
0.01$ from GRBs and $\xi_2 \gsim 10^{-9}$ from flaring AGN
[\cite{AGNs}]. Limits for higher values of $n$ are too small to be
of any relevance. Stronger bounds can be obtained under more
specific assumptions on the nature of LIV (see e.g.
[\cite{JLM03}-\cite{SS05}]).
We focus here on time delays as these provide the most model
independent LIV test [\cite{Ellis04}].

The time delay of a particle with energy, $E$, arriving from a
source at a distance $d$, is of order\footnote{See Eq.
\ref{LIDdelay} below for an exact formula.}:
\begin{equation}
\Delta t \approx  {1+n \over 2} \left ( {d \over c}\right ) \left
( {E \over \xi_n M_{pl}} \right )^n  \ . \label{deltat}
\end{equation}
To improve current limits one needs a more distant source, an
observation at higher energies or an improved temporal resolution
(provided that all particles are emitted simultaneously at the
source). However, pair production on the IR background limits the
distances that high energy photons can travel. The lower photon
number fluxes at higher energies limit, further,  the possible
time resolution [\cite{LIDGRB}]. Very high energy neutrino
[\cite{ck03}] provide an alternative that overcomes these
problems.

Practically all current GRB models (see  [\cite{GRBrev}] for a
review) predict bursts of very high energy neutrinos, with energy
ranging from 100TeV to 10${^4}$TeV (and possibly up to $10^6$TeV),
that should accompany GRB [\cite{WB,Vietri}]. As the energy of
these neutrinos are orders of magnitude higher than the energies
of photons observed from cosmological distances, the corresponding
time delays are longer and  can open a new window on the LIV
parameter space.

The 1637 bursts detected over an effective exposure time of 2.62
years and recorded in the BATSE 4B Catalog have an average fluence
of $1.2 \times 10^{-5}$ ergs/cm$^2$. Assuming that the emitted
neutrinos fluence is one tenth that observed in photons (a
reasonable assumption concerning the relevant interactions), we
obtain an average GRB induced $\nu$ flux of $5 \times 10^{18}$
eV/(km$^2$yr). Using the most likely value of $E_\nu \approx
100$TeV  and a detection probability of $10^{-4}$ in a km$^3$
detector, we estimate a detection rate of 5 events per year. This
rough estimate is in agreement with several model dependent
calculation [\cite{Waxman}-\cite{EventRates}]
that find a detection rate of  a few to a few dozen  events per
year. The increasing sensitivity of the detectors with the
neutrino energy [\cite{Guetta,Detectors}] compensates somewhat
over a decreasing flux (for a given fixed total emitted $\nu$
flux) and the detected flux decrease only like $E^{-0.5}$ for $E
\gsim 100$TeV. Thus, for Icecube only neutrinos up to $10^4$TeV
are relevant.

The detection probability of neutrinos from a given burst is
small\footnote{The detection probability  may reach 0.1 in an
extremely bright burst} $\sim 10^{-2}$. It is, therefore, unlikely
that two neutrinos will be detected from the same burst and a
direct comparison between the arrival time of two neutrino
[\cite{ck03}] cannot be done.   LIV measurements will have to
depend on the time delay between a single  detected neutrino and
the prompt low energy GRB photons.

The LIV time delay of a high energy neutrino with an observed
energy, $E$, emitted at redshift $z$ is\footnote{We safely neglect
the neutrino mass that adds, at these energies, an negligible
delay.}:
\begin{equation}\label{LIDdelay}
{\scriptstyle\Delta}t=\frac{1}{H_0}\int_0^z\left(\frac{1+n}{2}\left(\frac{E}{\xi
E_{pl}}\right)^n(1+z')^n\right)\frac{dz'}{\sqrt{\Omega_m(1+z')^3+\Omega_\Lambda}}
\ .
\end{equation}
Delays of order of hours are expected for a $100$TeV $\nu$ from a
$z=1$ burst with $\xi_1=1$ (or with $\xi_2=10^{-7}$). These values
should be compared with the time interval, $t_b(E,p)$,  in which
(at a given confidence level, $p$) no background neutrino from the
same direction in the sky is expected.

\begin{figure}[ht!]
   \centering
   \includegraphics[width=16cm,clip=true]{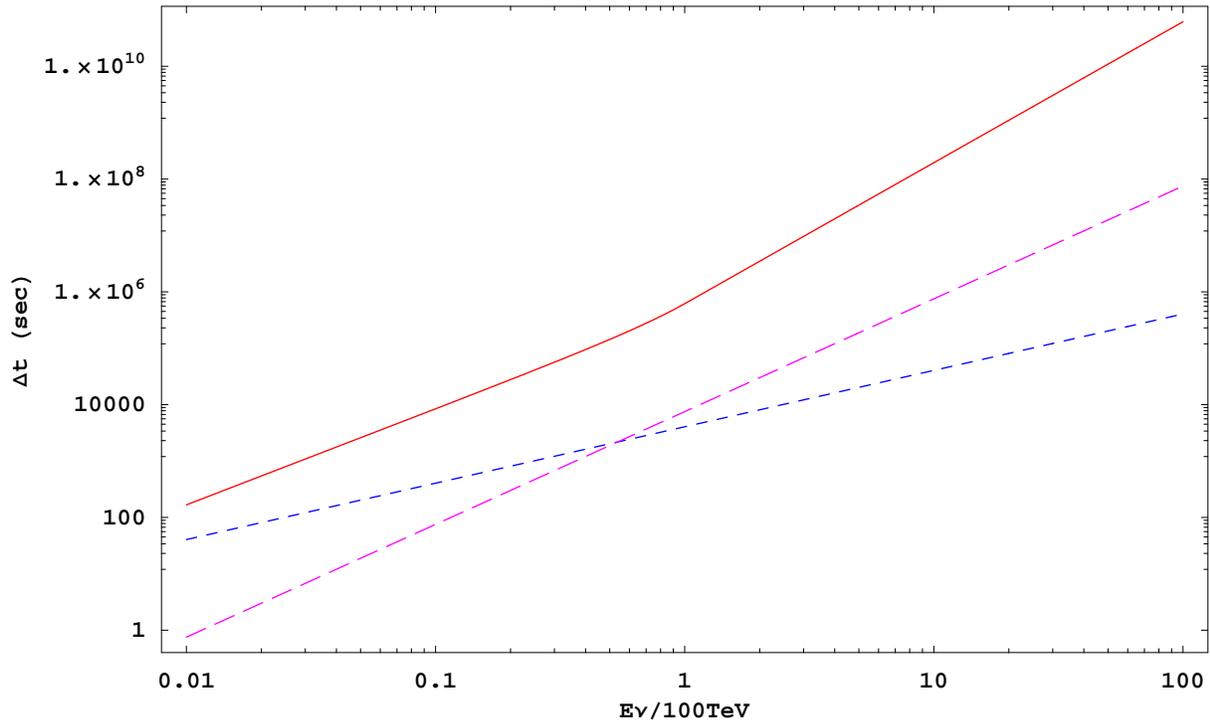}
   \caption{  $t_b(E,0.01)$ (solid red line) the time interval for
   $10^{-4}$ background events (corresponding to false positive
of 1\%) and the LIV time delays, $\Delta t$, from  $z=1$ and
$\xi_1=1$ (long dashed blue line) or $\xi_2=10^{-7}$ (short dashed
purple line).}
   \label{fig:DNn1}
\end{figure}

The dominant background arises  from muons produced by atmospheric
neutrinos. Using the atmospheric neutrino spectrum [\cite{Guetta}]
($\propto E^{-\beta}$) and the probability that a $\nu_\mu$
generates a detectable $\mu$ [\cite{Guetta}] ($\propto E^\alpha$)
we estimate the number of background events detected in a detector
of size A, from a solid angle $\Omega$ and during a time interval
$\Delta t$ as:
\begin{equation}
N_{bg}\simeq 5 \times 10^{-17} A\cdot\Omega \cdot \Delta t
\int_{E}^\infty { d \tilde E_\nu} {\tilde E_\nu}^{\alpha-\beta}
\end{equation}
where $\tilde E_\nu \equiv E_\nu/100$TeV, ($\alpha=1$,$\beta=3.7$)
for $E_\nu<100$TeV and ($\alpha=0.5$,$\beta=4$) for
$E_\nu>100$TeV. The currently constructed IceCube is designed to
determine the direction of muons with sub-degree accuracy
[\cite{Detectors}] corresponding to $\Omega \approx 10^{-3}$
square radians. Fig. \ref{fig:DNn1} depicts $t_b(E,0.01)$, the
interval in which we expect $10^{-4}$ background events
(corresponding to false alarm of 1\%).

An observed neutrino can be associated with a burst (and
interpreted as a positive detection of a time delay) if
$t_b(E,p)>{\scriptstyle\Delta}t$. As the detector is extremely
quiet at these energies a neutrino arriving months or even years
after the burst can be associated with the burst. Specifically,
for $\xi\gsim 1$ ($\xi_2 \gsim 10^{-7}$) the background does not
pose any problem. For n=1 one can explore using Grb $\nu$'s the
parameters up and above to the Planck scale, a region that cannot
be explored in any other way today [\cite{LIDGRB}].

\vskip 1cm
\begin{figure}[ht!]
   \centering
  \includegraphics[width=10cm,clip=true,angle=-90]{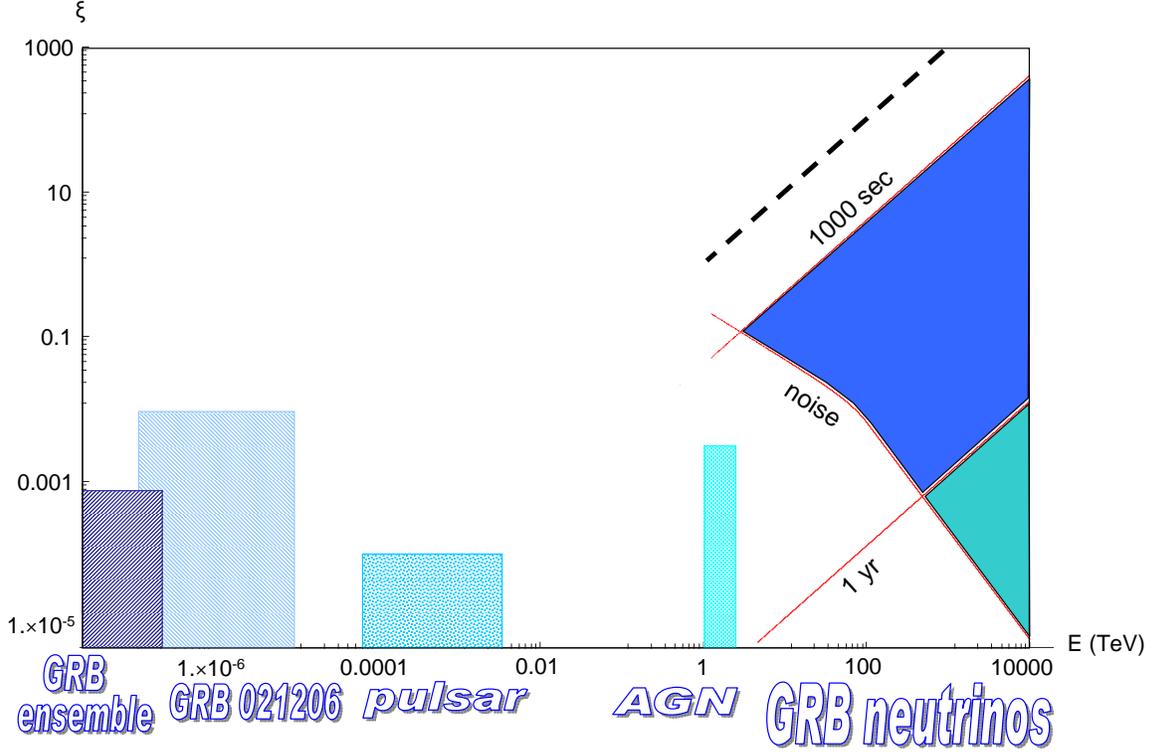}
   \caption{The range of  $\xi_1$ that can be explored for LIV delays  using
     GRB $\nu$'s
   from $z=1$ as a function of  $E_\nu$.  The colored regions
   indicate the range where $1000$sec$<\Delta
   t <  t_b(E,0.01)$. The additional condition $\Delta t <1$year is imposed in the
   upper (dark blue) range.  The dashed bold line describes the lower limit that can
be obtained from a simultaneous (within 30sec) detection of a high
energy neutrino and the prompt GRB. Also marked are current limit
obtained using various photon sources. }
   \label{fig:BOUNDSn1}
\end{figure}

\begin{figure}[ht!]
   \centering
   \includegraphics[width=10cm,clip=true,angle=-90]{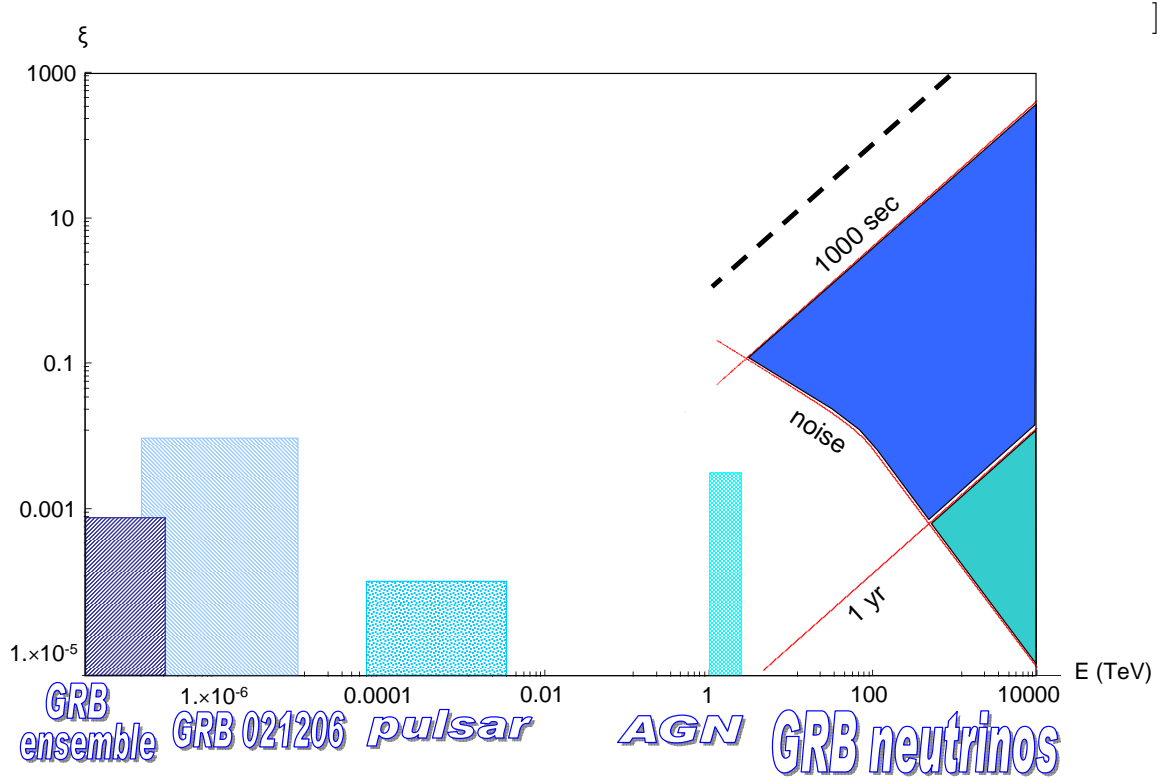}]
   \caption{The range of  $\xi_2$ that can be explored for LIV delays  using
    GRB $\nu$'s from $z=1$ as a function of  $E_\nu$. The colored regions
   indicate the range where $1000$sec$<\Delta
   t <  t_b(E,0.01)$. The additional condition $\Delta t <1$year is imposed in the
   upper (dark blue) range. The dashed bold line describes the lower limit that can
be obtained from a simultaneous (within 30sec) detection of a high
energy neutrino and the prompt GRB. Also marked are current limit
obtained using various photon sources. }
   \label{fig:BOUNDSn2}
\end{figure}

The range of LIV scale parameters, $\xi_n$, that can be detected
using LIV time delays of  GRB  $\nu$'s  is limited by additional
considerations. The duration of  long GRBs can be of order
$1000$sec and  it is not clear whether the $\nu$'s will accompany
the prompt emission or the early afterglow. We impose, therefore,
a conservative minimal delay of $1000$sec. Additionally we set an
arbitrary practical bound on  the maximal delay as a year. Figs.
\ref{fig:BOUNDSn1} and \ref{fig:BOUNDSn2} depict the region of
$\xi_{1,2}$ that can be determined by delayed detection of GRB
$\nu$s. This range is many orders of magnitude above current
limits. Detection of neutrinos coinciding with the prompt GRB
emission will set new upper limits to the LIV scale (see Figs.
\ref{fig:BOUNDSn1} and \ref{fig:BOUNDSn2}). No detection at all
will, of course, send us back to revise current GRB models.

Given the incomplete sky coverage of GRB detectors we expect a few
associations between a GRB and a $\nu$  event per year. Clearly no
matter what is the statistical significance a single detection
won't convince a skeptic observer. However, repeated detections
over several years of  $\nu$'s associated with GRBs with
compatible time delays and observed energies might do so.

This research was supported by a US-Israel BSF grant.

\end{document}